\documentclass{elsart}
\usepackage{epsfig}
\usepackage{enumerate}

\begin{document}
\begin{frontmatter}

\title{Phase transitions in the majority-vote model with two types of noises}

\author{Allan R. Vieira}
\thanks{allanrv@if.uff.br}
\author{and Nuno Crokidakis}
\thanks{nuno@if.uff.br}

\address{
Instituto de F\'{\i}sica, \hspace{1mm} Universidade Federal Fluminense \\
Av. Litor\^anea s/n, \hspace{1mm} 24210-340 \hspace{1mm} Niter\'oi - RJ, \hspace{1mm} Brazil}

\maketitle

\begin{abstract}
\noindent
In this work we study the majority-vote model with the presence of two distinc noises. The first one is the usual noise $q$, that represents the probability that a given agent follows the minority opinion of his/her social contacts. On the other hand, we consider the independent behavior, such that an agent can choose his/her own opinion $+1$ or $-1$ with equal probability, independent of the group's norm. We study the impact of the presence of such two kinds of stochastic driving in the phase transitions of the model, considering the mean field and the square lattice cases. Our results suggest that the model undergoes a nonequilibrium order-disorder phase transition even in the absence of the noise $q$, due to the independent behavior, but this transition may be suppressed. In addition, for both topologies analyzed, we verified that the transition is in the same universality class of the equilibrium Ising model, i.e., the critical exponents are not affected by the presence of the second noise, associated with independence.

\end{abstract}
\end{frontmatter}


\section{Introduction}

\qquad The study of dynamics of opinion formation is nowadays a hot topic in the Statistical Physics of Complex Systems, with a considerable amount of papers published in the last years (see \cite{galam_book,sen_book,pmco_book,rmp} and references therein). Even simple models can exhibit an interesting collective behavior that emerges from the microscopic interaction among indivuals or agents in a given social network. Usually those models exhibit nonequilibrium phase transition and rich critical phenomena, which justifies the interest of physicists in the study of opinion dynamics \cite{mjo,vilela_brady,diogo_brady,lima,vilela_brady2,nuno_pmco,nuno_bjp,lima2013,raimundo_pre,coreanos,wu_holme}.

The majority-vote model have been extensively studied in the last 20 years \cite{mjo}. In the standard model, each site of a square lattice has an Ising-spin variable whose state $\pm 1$ may be associated with the opinion of an individual in a given social community. The time evolution of the model is governed by an inflow dynamics, where the center spin is influenced by its nearest neighbors: an individual located at the central site adopts the minority sign of the spins in its neighborhood with probability $q$ and the majority sign with probability $1-q$. Numerical results indicate that the model undergoes a nonequilibrium phase transition at a critical noise $q_{c}\approx 0.075$, and the critical exponents were found to be the same as those of the equilibrium 2D Ising model, i.e., $\beta\approx 0.125$, $\gamma\approx 1.75$ and $\nu\approx 1.0$ \cite{mjo}. The transition separates an ordered phase, where one of the opinions dominates the population, from a disordered one, where the two opinions coexist in equal proportions. 

After that, many extensions of the original model were proposed. For example, we can highlight the interaction between two different classes of agents \cite{vilela_brady}, spins with three-state variables \cite{diogo_brady,lima}, noise distributed according to a bimodal distribution \cite{vilela_brady2}, diffusion \cite{nuno_pmco} and heterogeneous agents \cite{lima2013}. The influence of topology was also considered, with studies on random \cite{raimundo_pre}, hypercubic \cite{coreanos} and hyperbolic \cite{wu_holme} lattices, random graphs \cite{pereira_brady}, small-world \cite{campos,stone} and scale-free networks \cite{lima2006,chen}, among others. Some new universality classes were found for different topologies \cite{diogo_brady,lima2013,raimundo_pre,coreanos,wu_holme, pereira_brady,campos}, but in some of the above-mentioned modifications the critical exponents are Ising-like, i.e, they are the same as those of the original majority-vote model \cite{vilela_brady,lima,vilela_brady2,nuno_pmco}.

The consideration of the usual noise $q$ in the majority-vote model produces an effect similar to the introduction of the \textit{contrarians} in the population, individuals that adopt the choice opposite to the prevailing choice of the others, whatever this choice is \cite{galam_cont,lalama,galam_hung,andre_kuba,banisch,nuno_celia_victor}. Indeed, $q$ is the probability of an agent assume the opposite opinion shared by the local majority. The contrarian effect is called \textit{anticonformism} in the language of Social Sciences, and it is one kind of nonconformism \cite{willis,nail}. Another kind of nonconformist is \textit{independence}, where the agent also take cognizance of the group norm, but he/she decides to take one of the possible opinions ($\pm 1$ in the case of the majority-vote model) independently of the majority or the minority opinion in the group \cite{asch,jahoda}. Effects of independence on phase transitions in opinion models were considered recently \cite{sznajd_indep1,sznajd_indep2,sznajd_indep3,nuno_pla}.

In this work we consider the two mentioned kinds of nonconformity, anticonformity and independence. The last is introduced in the system as a probability $p$, in a way that a given agent chooses one of the two possible opinions with equal probability ($1/2$). Thus, the population evolves under the presence of two types of noises, represented by the parameters $q$ and $p$. We consider agents placed on fully-connected networks and on square lattices, and our interest is to study the effects of the two kinds of stochastic driving in the phase transitions of the model.

This work is organized as follows. In Section 2 we present the microscopic rules that define the model, and the analytical and numerical results are discussed in two distinct subsections, considering the mean-field approximation and the square lattice. Finally, our conclusions are presented in Section 3.


\section{Model and Results}

\qquad Our model is based on the majority-vote model \cite{mjo}. Every site of a given lattice with $N$ sites is occupied by an agent, and to each site $i$ we assign a random opinion $\sigma_{i}=\pm 1$ with equal probability $1/2$, corresponding to the two possible opinions in a certain subject. In the original model \cite{mjo}, a randomly chosen individual follows the opinion of the minority of its $4$ nearest neighbors with probability $q$ and adopts the majority sign of the spins in its neighborhood with probability $1-q$. In this work we will consider the effects of the independent behavior, where a randomly chosen individual acts independently of their neighbors with probability $p$. In this case, he/she chooses one of the two possible opinions $\pm 1$ with equal probability $1/2$. On the other hand, with the complementary probability $1-p$ we apply the original rule of the majority-vote model. Summarizing, considering the effects of the two noises $q$ and $p$, a given spin $\sigma_{i}$ is flipped with probability
\begin{equation}\label{eq1}
w_{i}=\frac{1}{2}\,(1-p)\,\left[1-\gamma\,\sigma_{i}\,S\left(\sum_{\delta}\sigma_{i+\delta}\right)\right] + \frac{p}{2} ~,
\end{equation}
\noindent
where $\gamma=1-2\,q$, $S(x)={\rm sgn}(x)$ if $x\neq 0$, $S(0)=0$ and the summation is over the nearest neighbors. Observe that in the absence of the independent behavior ($p=0$) we recover the standard flip probability of the majority-vote model \cite{mjo}.

Thus, our model presents two distinct types of stochastic driving, governed by the two noises $q$ and $p$. In the language of social sciences, we are considering two kinds of nonconformity, namely anticonformity (contrarian effect, parameter $q$) and independence (parameter $p$). We are interested in the critical behavior of the model. As the parameter $p$ is the novelty of the model, we will consider the quantities of interest as functions of $p$, for typical values of $q$. These quantities are magnetization per spin, the susceptibility and the Binder cumulant, given by 
\begin{eqnarray}\label{eq2}
m & = & \left\langle\frac{1}{N}\left|\sum_{i=1}^{N}\sigma_{i}\right|\right\rangle ~, \\ \label{eq3}
\chi & = & N\,(\langle m^{2}\rangle - \langle m\rangle^{2})  ~, \\ \label{eq4}  
U & = & 1 - \frac{\langle m^{4}\rangle}{3\,\langle m^{2}\rangle^{2}}  ~,
\end{eqnarray}
\noindent
respectively. In Eqs. (\ref{eq2}), (\ref{eq3}) and (\ref{eq4}), $\langle\, ...\, \rangle$ denotes a configurational average taken at steady states. 

In this work we will consider two distinct topologies for the model, the fully-connected network and the square lattice. These distinct cases will be treated separately in the following subsections.


\subsection{Fully-connected network}

\qquad In this section we consider that each agent can interact with all others, which corresponds to a mean-field limit. Following ref. \cite{maj_vot_mf}, one can derive analytically the behavior of the stationary magnetization $m$. A given configuration of the system can be denoted by $\{\sigma\}=(\sigma_{1},\sigma_{2},...,\sigma_{i},...,\sigma_{N})$. The time evolution of the probability $P(\{\sigma\},t)$ to found the system in the state $\{\sigma\}$ at a time $t$ is governed by the master equation \cite{vank,mjo_book}
\begin{equation} \label{eq5}
\frac{d}{dt}\,P(\{\sigma\},t) = \sum_{i}[w_{i}(\sigma^{i})\,P(\{\sigma^{i}\},t)-w_{i}(\sigma)\,P(\{\sigma\},t)] ~.
\end{equation}
\noindent
In the above equation, the state $\{\sigma^{i}\}$ can be obtained from the state $\{\sigma\}$ by the flip of the spin on site $i$, i.e., $\{\sigma^{i}\}=(\sigma_{1},\sigma_{2},...-\sigma_{i},...,\sigma_{N})$, and the factor $w_{i}(\sigma)$ can be interpreted as the flip rate of the $i$-th site ($\sigma_{i}\to-\sigma_{i}$), given by Eq. (\ref{eq1}). The summation is over all the sites. From Eq. (\ref{eq5}), one can get the time evolution of the average $\langle\sigma_{i}\rangle$,
\begin{equation} \label{eq6}
\frac{d}{dt}\,\langle\sigma_{i}\rangle = -2\langle\sigma_{i}\,w_{i}(\sigma)\rangle.
\end{equation}
Considering our modified transition rate given by Eq. (\ref{eq1}), Eq. (\ref{eq6}) gives us 
\begin{equation} \label{eq7}
\frac{1}{1-p}\,\frac{d}{dt}\,\langle\sigma_{i}\rangle = -\langle\sigma_{i}\rangle + \gamma\,\left\langle\,S\left(\sum_{\delta}\sigma_{i+\delta}\right)\right\rangle - \frac{p}{1-p}\,\langle\sigma_{i}\rangle ~.
\end{equation}
\noindent
In the mean-field limit, we choose a random site $i$, that will be the ``central'' site $\sigma_{i}$, and its $4$ ``nearest'' neighbors are also randomly chosen. If these neighbors are labeled as $\sigma_{1}$, $\sigma_{2}$, $\sigma_{3}$ and $\sigma_{4}$, one can write \cite{maj_vot_mf,mjo_book}
\begin{eqnarray} \nonumber
S\left(\sum_{\delta}\sigma_{i+\delta}\right) &  = & S(\sigma_{1}+\sigma_{2}+\sigma_{3}+\sigma_{4})=\frac{3}{8}(\sigma_{1}+\sigma_{2}+\sigma_{3}+\sigma_{4}) - \\\label{eq8}
&   &\mbox{}-\frac{1}{8}(\sigma_{1}\,\sigma_{2}\,\sigma_{3}+\sigma_{1}\,\sigma_{2}\,\sigma_{4}+\sigma_{1}\,\sigma_{3}\,\sigma_{4}+\sigma_{2}\,\sigma_{3}\,\sigma_{4}) ~.
\end{eqnarray}
\noindent
The mean-field approximation disregards all correlations, and one can write
\begin{eqnarray} \label{eq9}
\langle\sigma_{i}\rangle & = & m ~, \\  \label{eq10}
\langle\sigma_{i}\,\sigma_{j}\,\sigma_{k}\rangle & = & \langle\sigma_{i}\rangle\,\langle\sigma_{j}\rangle\,\langle\sigma_{k}\rangle = m^{3} ~.  
\end{eqnarray}
Thus, using Eqs. (\ref{eq8}), (\ref{eq9}) and (\ref{eq10}) in Eq. (\ref{eq7}), one obtains
\begin{eqnarray} \label{eq11}
\frac{1}{1-p}\,\frac{d}{dt}\,m = \left(-\epsilon\, - \frac{\gamma}{2}\,m^{2}\right)\,m ~,
\end{eqnarray}
\noindent
where 
\begin{equation} \label{eq12}
\epsilon=1-\frac{3\,\gamma}{2}+\frac{p}{1-p} ~.
\end{equation}

At the stationary state we have $(dm/dt)=0$, and Eq. (\ref{eq11}) give us two distinct solutions. The first one is the disordered state solution $m=0$, valid for $\epsilon>0$, and the second one is the ordered state solution $m=\sqrt{2\,|\epsilon|/\gamma}$, valid for $\epsilon<0$. The last solution, using the definition $\gamma=1-2\,q$ and Eq. (\ref{eq12}), gives us
\begin{eqnarray} \label{eq13}
m = \sqrt{\frac{2\,p/(1-p)+6\,q-1}{2\,q-1}}  ~,
\end{eqnarray}
\noindent
or in the usual form $m\sim (p-p_{c})^{\beta}$, where
\begin{equation} \label{eq14}
p_{c}=p_{c}(q)=\frac{6\,q-1}{6\,q-3}
\end{equation}

\begin{figure}[t]
\begin{center}
\vspace{6mm}
\includegraphics[width=0.4\textwidth,angle=270]{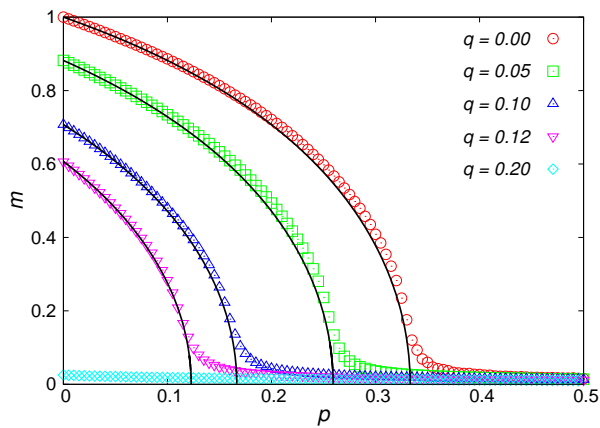}
\end{center}
\caption{(Color online) Order parameter $m$ as a function of the independence probability $p$ for typical values of the noise $q$, considering the mean-field formulation of the model. The symbols are numerical results for population size $N=10^{4}$, averaged over $100$ independent simulations, and the full lines are the analytical prediction given by Eq. (\ref{eq13}). Observe that for sufficiently large values of $p$ and $q$ the system is disordered, which corresponds to the analytical solution $m=0$.}
\label{fig1}
\end{figure}

\noindent
and we found a typical mean-field exponent $\beta=1/2$, as expected, indicating that the model should belong to the mean-field Ising universality class. Eq. (\ref{eq14}) give us $q_{c}=1/6$ for $p=0$, in agreement with the standard mean-field majority-vote model \cite{maj_vot_mf}, i.e., in the absence of independence. In addition, this result shows that there is another transition due to independence, $p_{c}(q=0)=1/3$, i.e., even in the absence of the usual noise $q$ the system undergoes an order-disorder transition due to the presence of the independent behavior. Summarizing, there is an order-disorder phase transition in the plane $p$ versus $q$ for $0\leq\,p<1/3$ and $0\leq\,q<1/6$; otherwise, the only valid solution is the disordered one, $m=0$. In other words, for sufficient large independence probability $p$ ($>1/3$), the phase transition is suppressed, and the system is disordered for all values of $q$. Finally, another important result comes from Eq. (\ref{eq13}): the consensus states with $m=1$ are obtained only for $p=q=0$, in the absence of the two noises.

To verify the above calculations, we performed computer simulations of the majority-vote model considering a fully-connected graph. In this case, all the $5$ agents considering during an interaction are randomly chosen. In Fig. \ref{fig1} we show the magnetization per spin as a function of $p$, estimated numerically from Eq. (\ref{eq2}) and a comparison with the derived Eq. (\ref{eq13}), for typical values of $q$. Notice that for $q>1/6$ there is no transition anymore, for all values of $p$, in agreement with the analytical prediction that give us $m=0$ (see the curve for $q=0.20$ in Fig. \ref{fig1}). In addition, one can see a transition even in the absence of the usual noise $q$, generated by the independent behavior of the agents.

\begin{figure}[t]
\begin{center}
\vspace{6mm}
\includegraphics[width=0.4\textwidth,angle=270]{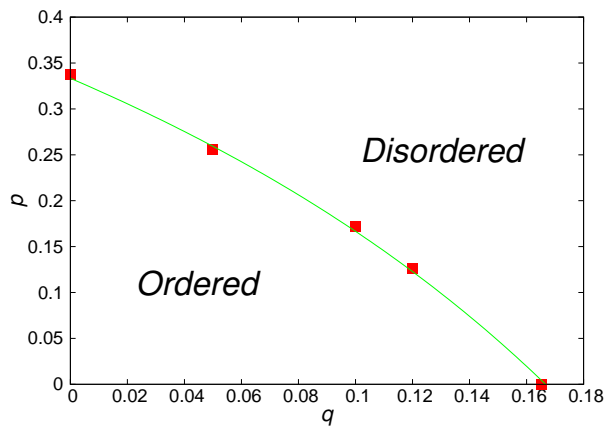}
\end{center}
\caption{(Color online) Phase diagram of the mean-field formulation of the model, in the plane $p$ (independence probability) versus $q$ (usual noise). One can see the Ordered and the Disordered phases. The squares are the numerical estimates of the critical points $p_{c}(q)$, and the full line is the analytical prediction, Eq. (\ref{eq14}). The error bars are smaller than data points.}
\label{fig2}
\end{figure}

As a final discussion, we performed a finite-size scaling (FSS) analysis (not shown) for typical values of $q<q_{c}=1/6$ in order to obtain numerical estimates for the critical exponents of the model, as well as for the critical points $p_{c}(q)$. Based on the standard FSS equations, 
\begin{eqnarray} \label{eq15}
m(N) & \sim & N^{-\beta/\nu} \\  \label{eq16}
\chi(N) & \sim & N^{\gamma/\nu} \\   \label{eq17}
U(N) & \sim & {\rm constant} \\   \label{eq18}
p_{c}(N) - p_{c} & \sim & N^{-1/\nu} ~,
\end{eqnarray}
we obtained $\beta\approx 1/2$, $\gamma\approx 1$ and $\nu\approx 2$, for all values of $q$, indicating that the model belongs to the mean-field Ising universality class, as suggested by the analytical calculations. In addition, the FSS analysis allows us to identify the critical points $p_{c}(q)$ by the crossing of the Binder cumulant curves for distinct population sizes $N$. Considering the estimated values of $p_{c}(q)$ for typical values of $q$, we exhibit in Fig. \ref{fig2} the phase diagram of the model in the plane $p$ versus $q$. We also plot the analytical prediction, Eq. (\ref{eq14}), and one can see that we have a good agreement among the analytical and numerical results.


\subsection{2D square lattice}

\qquad In this subsection we simulate our model on square lattices of distinct sizes $L$, in order to verify if the mean-field predictions are at least qualitatively valid when we consider a neighborhood. 

\begin{figure}[t]
\begin{center}
\vspace{6mm}
\includegraphics[width=0.4\textwidth,angle=270]{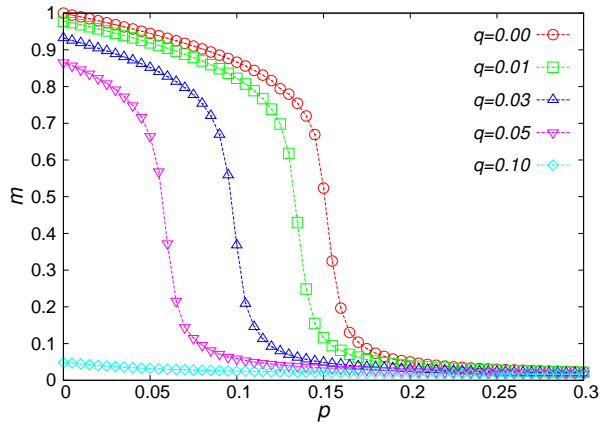}
\end{center}
\caption{(Color online) Order parameter $m$ as a function of the independence probability $p$ for typical values of the noise $q$, considering the model defined on a 2D square lattice of linear size $L=100$. Observe that for sufficiently large values of $p$ and $q$ the system is disordered. Each point is averaged over $100$ independent simulations, and the lines are just guides to the eye.}
\label{fig3}
\end{figure}

In Fig. \ref{fig3} we exhibit results for the magnetization per spin as a function of $p$, for $L=100$ and typical values of the noise $q$. One can observe an order-disorder phase transition even for $q=0.0$, and for sufficiently large values of $q$ there is no transition anymore. This result is in agreement with the numerical estimate of the critical point $q_{c}(p=0.0)\approx 0.075$ \cite{mjo}. As in the previous case, for sufficient large values of $p$ ($>\approx 0.15$, as we will see in the following), the phase transition is suppressed, and the system is disordered for all values of $q$.

\begin{figure}[t]
\begin{center}
\vspace{0.5cm}
\includegraphics[width=0.3\textwidth,angle=270]{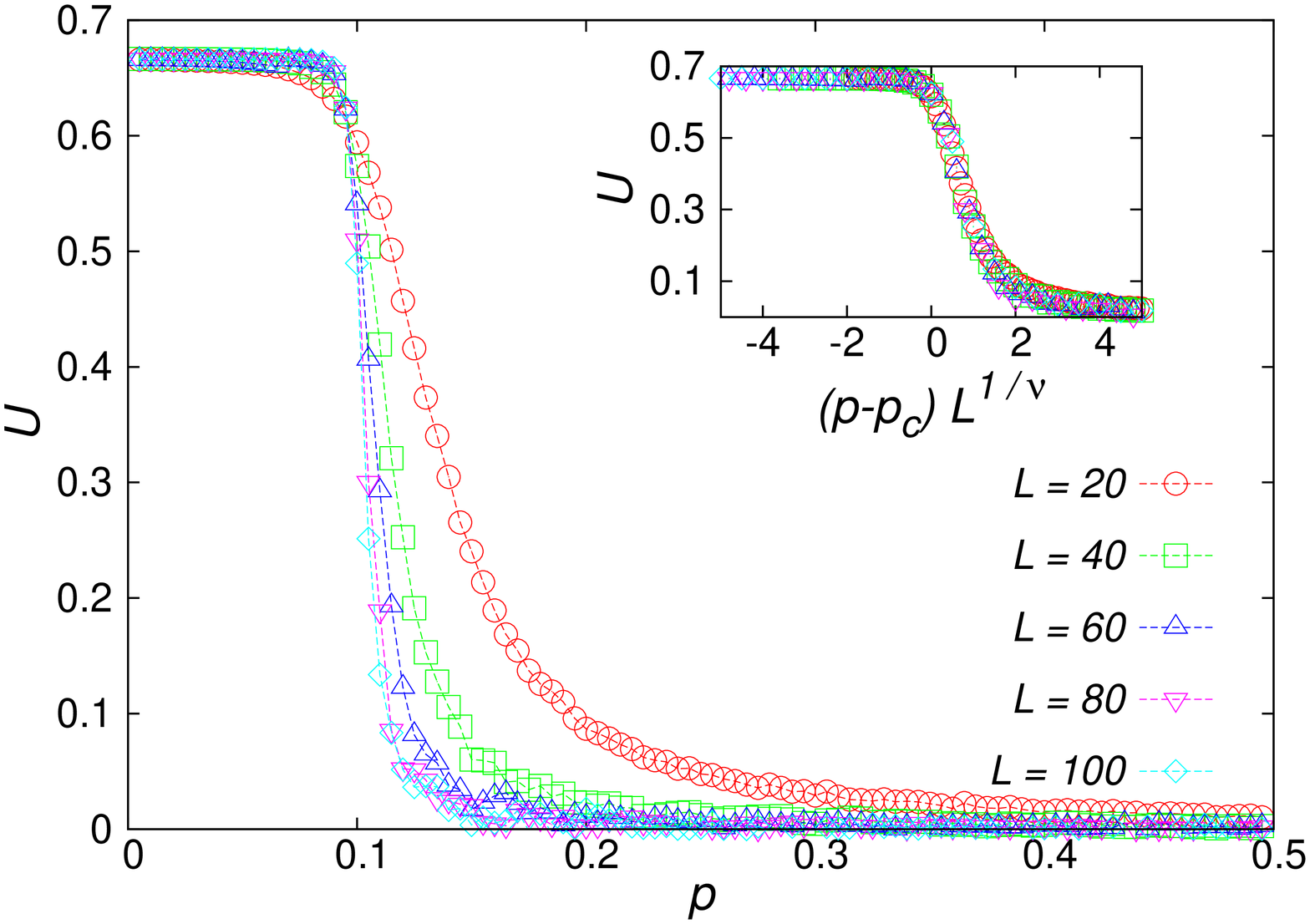}
\hspace{0.5cm}
\includegraphics[width=0.3\textwidth,angle=270]{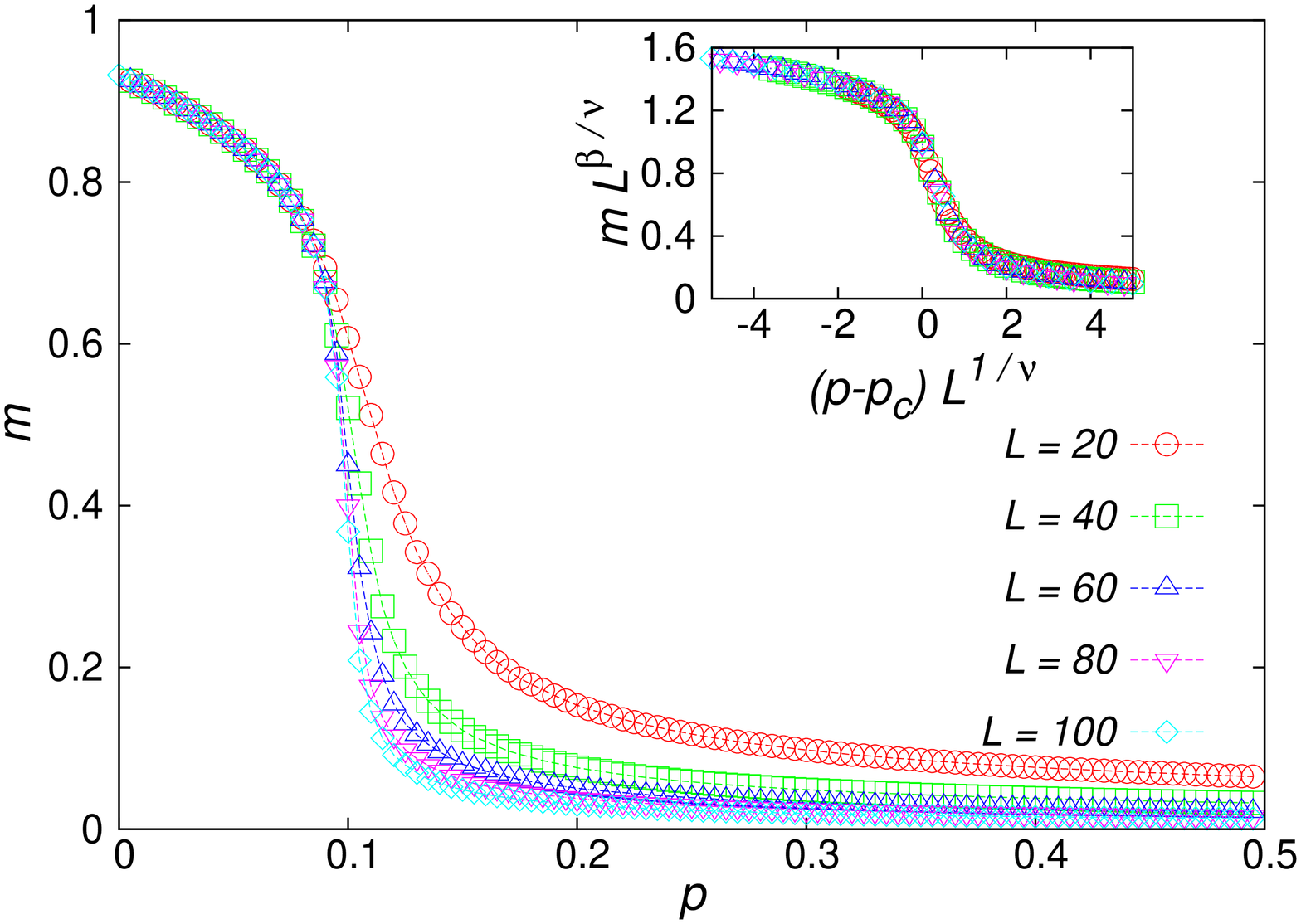}
\\
\vspace{1.0cm}
\includegraphics[width=0.3\textwidth,angle=270]{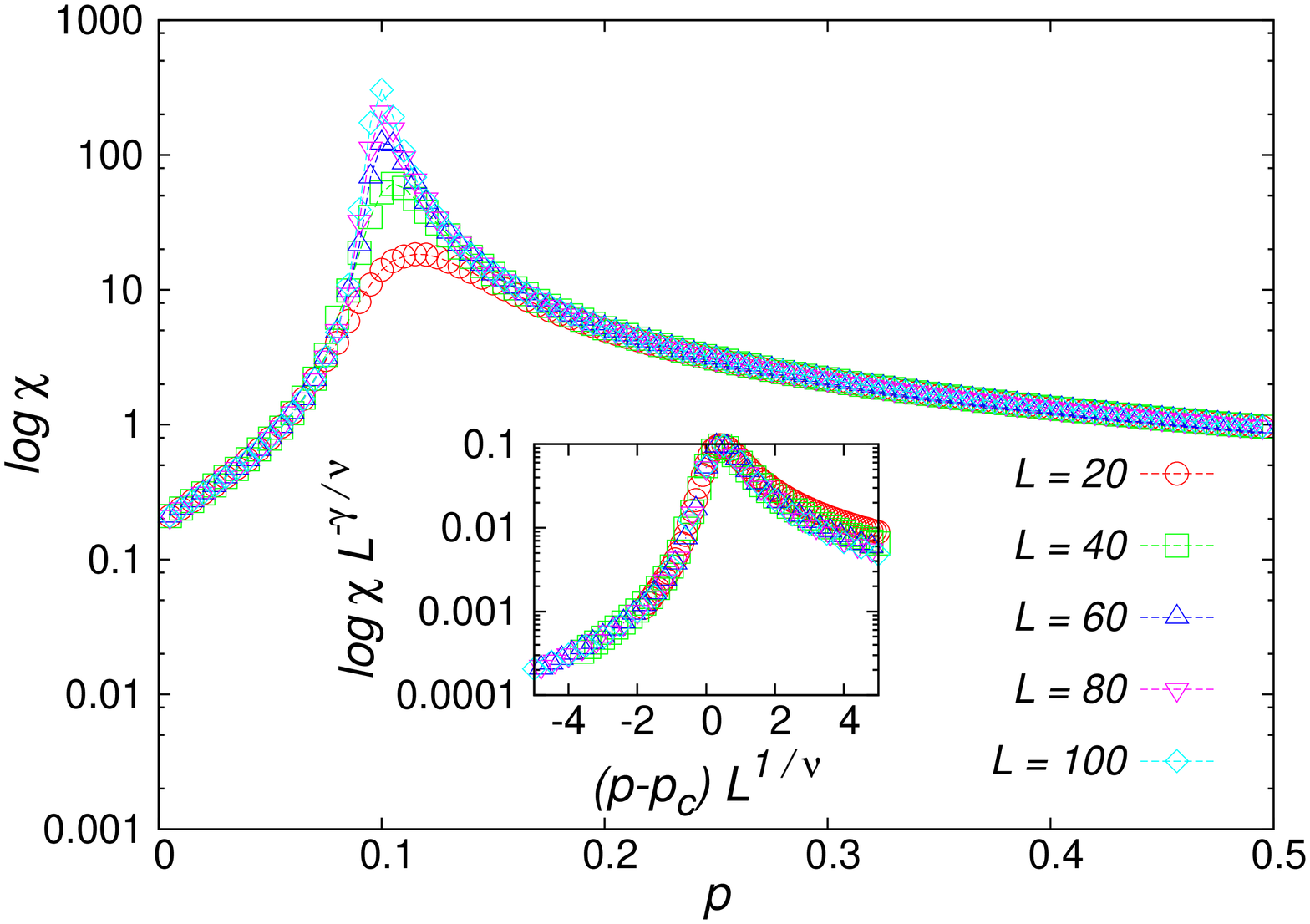}
\end{center}
\caption{(Color online) Finite-size scaling analysis of the transition for the model on a 2D square lattice, for $q=0.03$. The best collapse of data was found for $p_{c}\approx 0.095 $, $\beta\approx 1/8$, $\gamma\approx 7/4$ and $\nu\approx 1$.}
\label{fig4}
\end{figure}

One can estimate the critical points and exponents considering the FSS Eqs. (\ref{eq15}) to (\ref{eq18}), considering the change $N\to L$. One example is exhibited in Fig. \ref{fig4}, for $q=0.03$. We obtained the same critical exponents for all values of $q<q_{c}(0.0)$, $\beta\approx 1/8$, $\gamma\approx 7/4$ and $\nu\approx 1$, i.e., the model on the square lattice belongs to the 2D Ising model universality class. In other words, the critical exponents are not affected by the introduction of another noise in the model, namely independence.

Thus, the mean-field formulation presented in the previous subsection predicts some important behaviors: (i) there is an order-disorder transition even for $q=0.0$, generated by the independent behavior; (ii) the model is in the Ising model universality class; (iii) the consensus states $m=1$ are obtained only for $p=q=0.0$. Despite the quantitative differences between the two formulations, mean field and square lattice, the above results are in agreement with the simulations of the model on the square lattice.

\begin{figure}[t]
\begin{center}
\vspace{6mm}
\includegraphics[width=0.4\textwidth,angle=270]{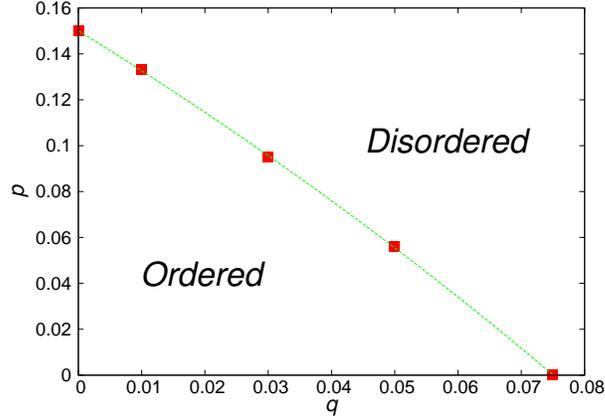}
\end{center}
\caption{(Color online) Phase diagram of the model defined on a 2D square lattice, in the plane $p$ versus $q$, separating the Ordered and the Disordered phases.  The squares are the numerical estimates of the critical points $p_{c}(q)$, obtained by the crossing of the Binder cumulant curves. The error bars are smaller than data points, and the dashed line is a qualitative description of the phase boundary, as discussed in the text.}
\label{fig5}
\end{figure}

Considering the critical points $p_{c}(q)$ estimated from the crossing of the Binder cumulant curves, we exhibit in Fig. \ref{fig5} the phase diagram of the model on the square lattice, in the plane $p$ versus $q$. Based on Eq. (\ref{eq14}), we propose a qualitative description of the phase boundary that separates the ordered and the disordered phase, based on the relation
\begin{equation} \label{eq19}
p_{c}=p_{c}(q)=\frac{a\,q+b}{c\,q+d} ~,
\end{equation}
\noindent
with 4 parameters $a$, $b$, $c$ and $d$. Fitting data, we obtained the estimates $(a,b,c,d) \simeq (82.76,-6.21,80.54,-41,39)$. This heuristic description presents a good agreement with the critical points obtained from FSS analysis, as shown in Fig. \ref{fig5}. In particular, for $q=0$ we have $p_{c}(0)=(b/d)\approx 0.15$ and $q_{c}=(-b/a)\approx 0.075$ for $p=0$, and the last result is in agreement with the result obtained in the original majority-vote model \cite{mjo}.

\section{Final remarks}   

\qquad In this work, we have studied the majority-vote model in the presence of two noises. The first one is the usual noise $q$, representing the hesitance of an agent to follow a local majority opinion. In this case, this behavior is similar to the Galam's contrarians \cite{galam_cont}. On the other hand, we considered the independent behavior, governed by a probability $p$ that an agent chooses his/her own opinion $+1$ or $-1$ with equal probability, independent of the local majority or minority opinion. We considered the model on a fully-connected network (mean-field formulation) and on a square lattice, and despite the quantitative differences between the two formulations, as expected, the results in both cases are in qualitative agreement.
 
We performed Monte Carlo simulations of the model for distinct population sizes. In this case, the above-mentioned two noises contribute to disorder the system. For both cases (mean field and 2D), consensus states are only obtained in the total absence of noise, i.e., for $p=q=0.0$. On the other hand, we have an order-disorder transition even in the absence of the usual noise $q$, due to presence of the independent behavior in the population. Furthermore, the phase transition is in the same universality class of the equilibrium Ising model, i.e., the critical exponents are not affected by the presence of a second noise (independence). This result is in accordance to the Grinstein's criterion that states that nonequilibrium stochastic spin-like systems with up-down symmetry in regular lattices fall into the same universality class of the equilibrium Ising model \cite{mjo,grinstein,baxter_book}. At mean-field level, the results were complemented by analytical calculations, based on the master equation.

We also verified that there is a threshold value $p_{c}(q=0.0)$ above which the system is disordered for all values of $q$. In terms of a public debate with two distinct choices, it means that if at least a fraction $p_{c}(q=0.0)$ of the population takes independent decisions, there is no final decision in the debate, independent of the presence of contrarians. Otherwise, there can be a majority opinion dominating the population, which means a decision on the debate. In the mean-field approximation of the model, we showed analytically that the mentioned threshold value is $p_{c}(q=0.0)=1/3$.

As an extension of this work, it would be interesting to verify how results change with the number of neighbors. It could occur, similarly as in the q-voter model \cite{sznajd_indep2}, that the phase transition changes its nature from continuous to discontinuous with the increase of the number of neighbors.

\section*{Acknowledgments}

The authors acknowledge financial support from the Brazilian funding agency CNPq.

\end{document}